\documentclass{article}
\usepackage{multirow}
\usepackage{spconf,amsmath,epsfig}
\usepackage{subcaption}
\usepackage{color}
\let\OLDthebibliography\thebibliography
\renewcommand\thebibliography[1]{
  \OLDthebibliography{#1}
  \setlength{\parskip}{0pt}
  \setlength{\itemsep}{0pt plus 0.3ex}
}

\begin{document}\sloppy

\def\x{{\mathbf x}}
\def\L{{\cal L}}

\title{XGC-VQA: A unified video quality assessment model for User, Professionally, and Occupationally-Generated Content}
%
\name{Xinhui Huang$\tiny{^{1}}$*\thanks{*These authors contributed equally to this work.}, Chunyi Li$\tiny{^{1}}$*, Abdelhak Bentaleb$\tiny{^{3}}$, Roger Zimmermann$\tiny{^{2}}$,and Guangtao Zhai$\tiny{^{1}}$ \vspace*{-3mm}}
\address{Shanghai Jiao Tong University\textsuperscript{1}, National University of Singapore\textsuperscript{2}, Concordia University\textsuperscript{3} \vspace*{-4mm}}

\maketitle

\begin{abstract}
With the rapid growth of Internet video data amounts and types, a unified Video Quality Assessment (VQA) is needed to inspire video communication with perceptual quality. To meet the real-time and universal requirements in providing such inspiration, this study proposes a VQA model from a classification of User Generated Content (UGC), Professionally Generated Content (PGC), and Occupationally Generated Content (OGC). In the time domain, this study utilizes non-uniform sampling, as each content type has varying temporal importance based on its perceptual quality. In the spatial domain, centralized downsampling is performed before the VQA process by utilizing a patch splicing/sampling mechanism to lower complexity for real-time assessment. The experimental results demonstrate that the proposed method achieves a median correlation of $0.7$ while limiting the computation time below 5s for three content types, which ensures that the communication experience of UGC, PGC, and OGC can be optimized altogether.

\end{abstract}
\section{Introduction}
\label{sec:intro}

Video has become the dominant data type in today's internet and accounts for 82\% of network bandwidth usage~\cite{cisco2020cisco}. To cope with the massive amount and different types of video data being transmitted, classical Video Quality Assessment (VQA) has been used as an evaluation criterion for video transmission or encoding performance~\cite{intro-performance_guidance}. However, with the rapid development of perception-inspired communication~\cite{intro-perception}, VQA has been increasingly used to inspire video communication beyond just an overall perceptual quality gauge, such as network resource allocation\cite{inspire3}, video coding mode selection\cite{inspire2}, and real-time bitrate guidance~\cite{intro-realtime}. Due to the real-time requirement of services above, a unified VQA metric for various video contents is needed in a No Reference (NR) scenario with low-complexity.

However, the differences across video content types create a great challenge for designing such a unified model. For today's mainstream video providers, User Generated Content (UGC), Professionally Generated Content (PGC), and Occupationally Generated Content (OGC) are the three major content types. UGC~\cite{intro-platform} is created by a regular user of social platforms, PGC ~\cite{intro-PGC} is quality content created by professional users, and OGC~\cite{database-ugcpgcogc} is produced by practitioners. 

When offering perception-inspired bitrate guidance in media delivery systems, evaluating the perceptual quality of all video frames would compromise real-time performance, so downsampling is required. All three video types require downsampling in both the spatial and temporal domains. Due to the huge differences in resolution, luminance, and quality between UGC, PGC, and OGC, it is not feasible to use the same VQA method of downsampling for all three. For example, there exist already some fast and remarkable VQA metrics~\cite{intro-ogcmetric1, intro-ogcmetric2}, but they do not perform well on UGC. 
There exist also low-complexity metrics to deal with the challenge of UGC VQA, but their results on OGC are not consistent with the Human Visual System (HVS) due to OGC's higher resolution and dynamic range. Considering the differences between UGC, PGC, and OGC, how to build a unified VQA metric is still an open research question.

\vspace{-2mm}
\section{Related Works}
\label{sec:relate}
In the spatial domain, classic VQA models use a visual saliency map for downsampling. Xu \emph{et al.}~proposed to learn the video saliency model about the state-of-the-art H.265 codec features~\cite{r-spatial-map}. Another metric is to randomly sample video fragments into patches~\cite{r-patch}. However, since the HVS has different visual saliency for UGC, PGC, and OGC, their respective saliency maps are also different. Therefore, random sampling has difficulty to obtain a high-saliency region. On the other hand, calculating the saliency map will introduce extra time complexity, which contradicts the real-time requirement. Thus a simple and efficient method for sampling the space domain is needed.

\begin{figure*}[t]  
\centering 
\includegraphics[width=0.9\textwidth]{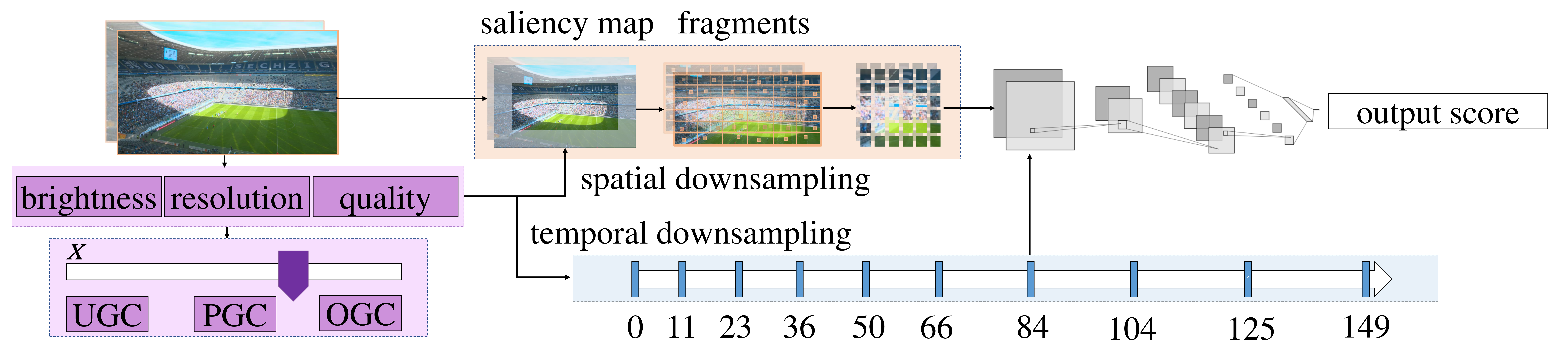}
\caption{The framework of the proposed method.} 
\label{fig:framework}
\vspace{-3mm}
\end{figure*}

In the temporal domain, the traditional approach is to consider several continuous frames~\cite{fastvqa}. However, this can lose a lot of temporal information. Another method is to subsample every few frames evenly~\cite{min2020study}, resulting in insufficiency of long-term features~\cite{yan2022subjective}. However, for UGC, users tend to switch to the next video after the first few seconds when watching low-quality videos. Conversely, for OGC, where the quality of the video is higher, users are more likely to watch the whole video and remember its later parts better, so the subsequent frames are more important.

\begin{figure}[t]  
\centering 
\includegraphics[width=0.4\textwidth]{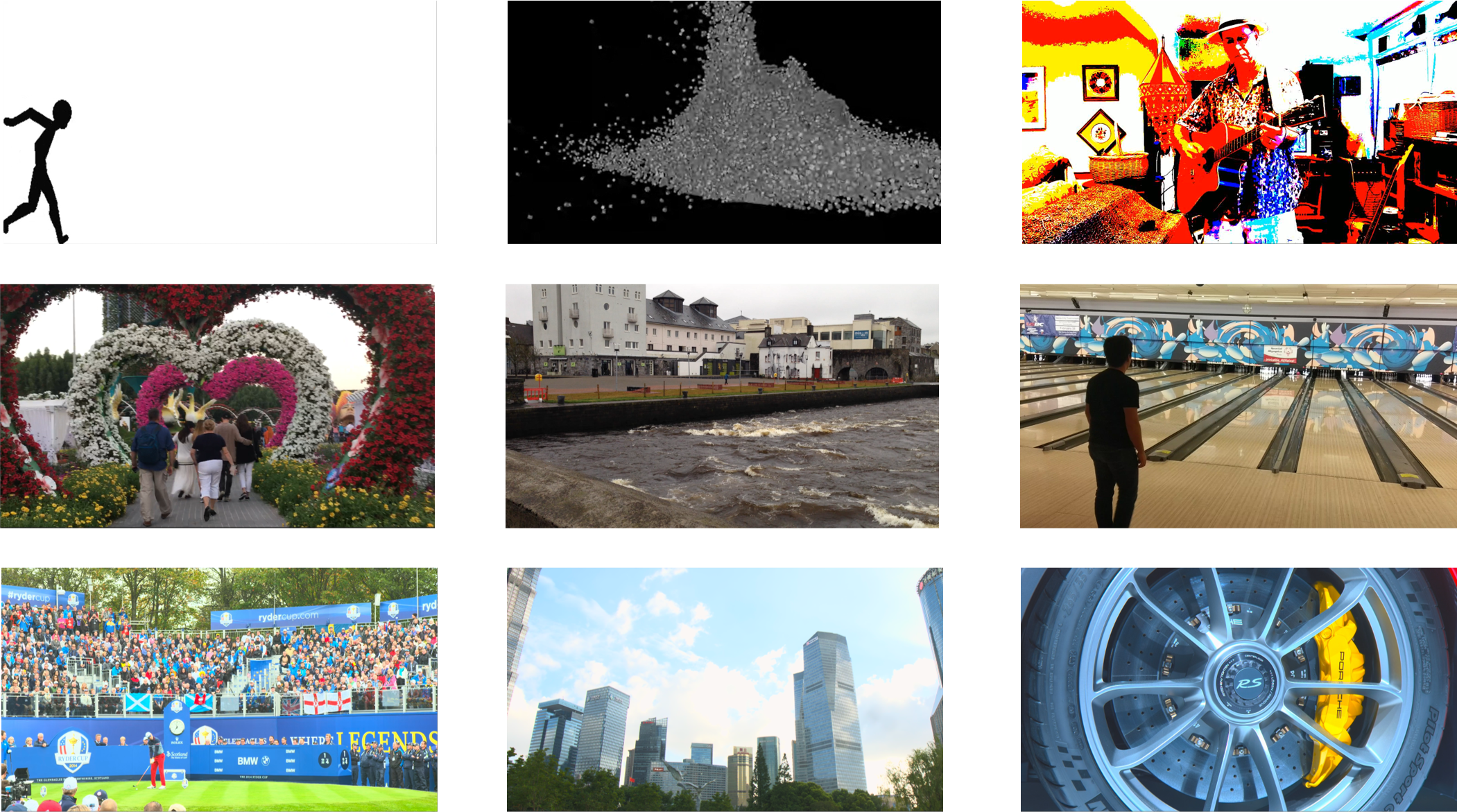}
\caption{Example of illustrating the difference UGC, PGC, and OGC: from the first to the third row, UGC, PGC, and OGC are listed in order of increasing resolution from left to right.}  
\label{fig:res}
\vspace{-4mm}
\end{figure}

Due to the limited generalization ability of the above model in UGC, PGC, and OGC, some databases\cite{database-ugcpgcogc} and metrics\cite{unified} are designed for this unified VQA task. However, with the development of internet video services, the specific content of UGC and OGC has also changed. On one hand, as UGC is easier to produce\cite{cisco2020cisco} in recent years, its overall quality has generally declined; on the other hand, with the application of High Dynamic Range (HDR) ~\cite{util-HDR1,util-HDR2} and the standardization of display devices\cite{sun2020dynamic}, the dynamic range of high-quality OGC increased, and some minority resolutions (e.g. 960*540) no longer appear in OGC. Thus, those unified databases / VQA methods should adapt today's video content.

Based on these insights, we design XGC VQA, where X stands for the attribute user, professional or occupational, with contributions in the following three aspects. (\emph{i}) \textbf{Classification}: We introduce an effective classification model for UGC, PGC, and OGC video through a parameter that is used to define the content producer's professionalism. This allows for different downsampling mechanisms for different content.
(\emph{ii}) 
\textbf{Spatial domain:} A centralized downsampling before the VQA process is conducted based on the patch splicing/sampling mechanism in FastVQA~\cite{fastvqa}. The sampling density depends on the above professionalism, thus minimizing the input for each UGC, PGC, and OGC without affecting the performance of the model.
(\emph{iii}) \textbf{Temporal domain:} Non-uniform sampling based on different temporal frame importance in UGC, PGC, and OGC. Sampling according to importance allows further reduction in model complexity without compromising performance.

\vspace{-2mm}
\section{Proposed Method}
\vspace{-1mm}

When aiming to apply a real-time unified VQA metric to all UGC, PGC, and OGC videos, we need to classify a video first and adopt different downsampling strategies for their specific content. The confidence parameter $x$ for XGC is obtained by a linear combination of the video features. In the spatial domain, we choose different attention maps according to the video types; in the temporal domain, we choose the frames to be evaluated according to the confidence parameter. The framework of our model is shown in Fig.~\ref{fig:framework}.

\subsection{Classification Modules}
\label{sec:classification}
Among the nine features of the video according to the previous study~\cite{database-ugcpgcogc}, the greatest differences among UGC, PGC, and OGC are brightness, resolution, and image quality. In this study, we first separate UGC from PGC, then distinguish between PGC and OGC.

The difference between UGC and PGC is that UGC is recorded by ordinary users and is not as professional as PGC in terms of equipment, and this inferiority is mainly reflected in two hardware-level constraints as follows.
(\emph{i}) \textbf{Brightness:}
    UGC is often shot with mobile phones, which can result in poorly lit footage due to the limitations of the device's camera and lighting conditions. Additionally, shooting with the front-facing camera can result in uneven brightness due to the camera's placement and the lighting direction~\cite{class-brightness}.
 (\emph{ii}) \textbf{Resolution:}
   UGC camera clarity is not high, or limited by network bandwidth, storage space, etc. The resolution of UGC is often between 360p--720p, while PGC and OGC can both reach 1080p or even 4K \cite {database-UGC, database-VQC, database-HDR}.

Fig.~\ref{fig:res} shows several selected screenshots from the UGC, PGC, and OGC video categories \cite {database-UGC, database-VQC, database-HDR}, and we note that the PGC photo has a higher resolution and more uniform luminance distribution, leading to a better experience for the user.

\begin{figure}[t]  
\centering 
\includegraphics[width=0.4\textwidth]{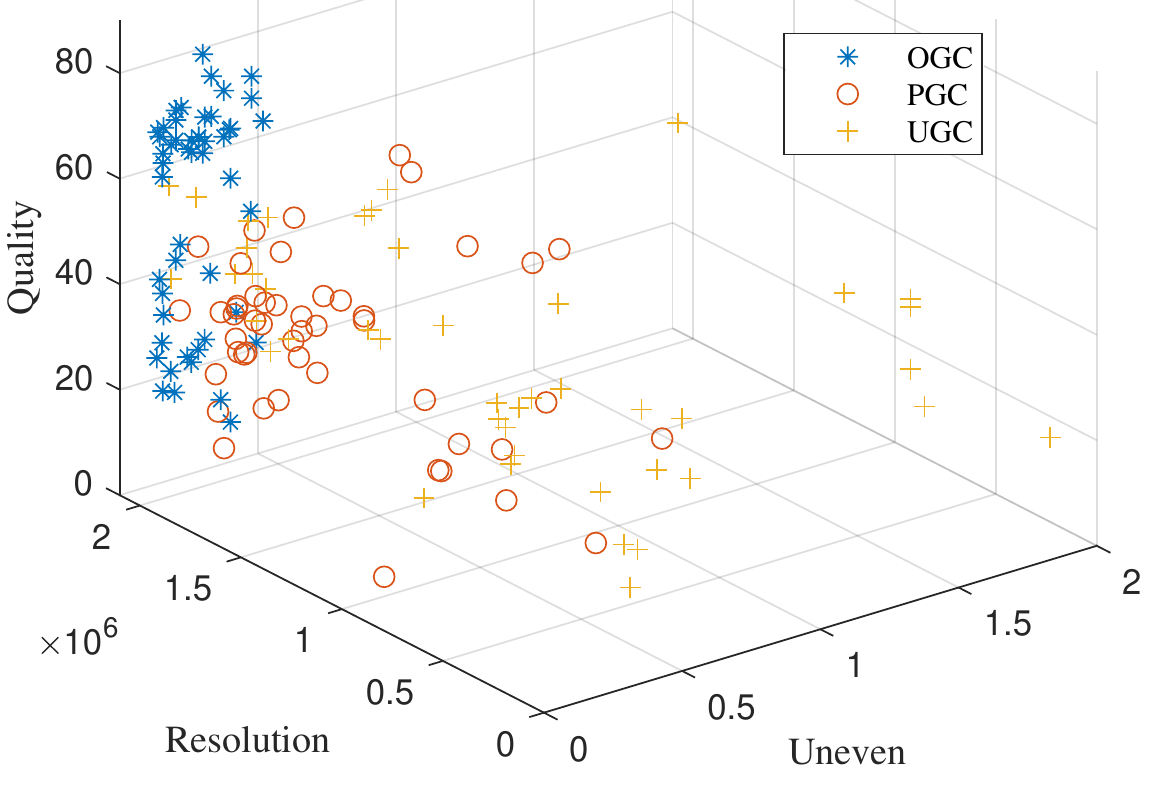}
\caption{Feature distribution of UGC, PGC, and OGC.} 
\label{fig:point}
\vspace{-3mm}
\end{figure}

Therefore, when we distinguish between UGC and PGC, we mainly consider these two aspects. Generally speaking, UGC has at least one defect in unevenness or resolution, and when both are good then UGC can evolve into PGC, so we use the worst of the two to characterize how much a UGC tends to be PGC, the hardware performance $\lambda$ and the confidence level $x$ can be expressed as:
\vspace{-2mm}
\begin{equation}
\lambda  = \min (1 - \frac{{{\rm std(img)}}}{{\rm mean(img)}},\frac{{\sqrt {h \cdot w} }}{{\sqrt {h_m \cdot w_m} }})
\label{edu:uneven}
\end{equation}
\vspace{-3mm}
\begin{equation}
x = \alpha \lambda \quad {\rm{if~}} ( \lambda  \le {\rm{1}})
\end{equation}
where $\alpha$ is a linearity coefficient. Empirically we found the best results when $\alpha$ equals 0.5. $img$ represents a frame in the video being evaluated, and $(h,w)$ indicate the height and width of the video. $(h_m,w_m)$ is an empirically\cite{database-UGC} pre-defined UGC resolution bound for videos, whose resolution below that tends to be classified as UGC. When $\lambda<1$, it means that hardware (uneven, unclear) is the limiting factor for a good video. Thus we have a valid distinction between UGC and PGC.

When $\lambda>1$, it means that the capture impairment and network are already good enough for evenness and resolution, and at this point, using better lenses or improving resolution, the video quality may not be improved because the monitor at the receiving end is not good enough~\cite{class-monitor}, or the Just Noticeable Difference (JND) of HVS is not triggered~\cite{class-JND}. The difference between PGC and OGC videos at this point is the quality of the content. 
Fig.~\ref{fig:res} shows the difference between PGC and OGC, and we can see that OGC is more focused on aesthetic quality and gives a better experience. 

The video quality relies heavily on deep learning \cite{class-quality}. Since we are a real-time model, the complexity introduced by multiple pooling convolutions of the neural network is unacceptable. Given that distortion quality is strongly correlated to aesthetic~\cite{yang2020fidelity}, we will use the commonly used distortion model~\cite{brisque} to characterize quality, and the confidence level $x$ can be expressed as follows:
\vspace{-1mm}
\begin{equation}
\begin{aligned}
    x=\alpha+(1-\alpha) \cdot \frac{{\rm brisque}(img)}{100} & \quad \text{  if $(\lambda > 1)$}
\end{aligned}
\end{equation}
where $\rm brisque(\cdot)$ represents the most widely used NR quality model~\cite{brisque}. This allows us to make a valid distinction between UGC, PGC, and OGC.

Fig.~\ref{fig:point} shows the luminance unevenness, resolution, and quality distribution in the UGC\cite{database-UGC}, PGC\cite{database-VQC}, and OGC\cite{database-HDR} database. We see that U/PGC and P/OGC have certain intersections, while U/OGC has almost no intersection, which shows that our classification method is accurate and effective.

\subsection{Spatial Domain}

\begin{figure}[tbp]
	\centering
	\begin{minipage}[t]{0.15\textwidth}
		\includegraphics[width=\textwidth]{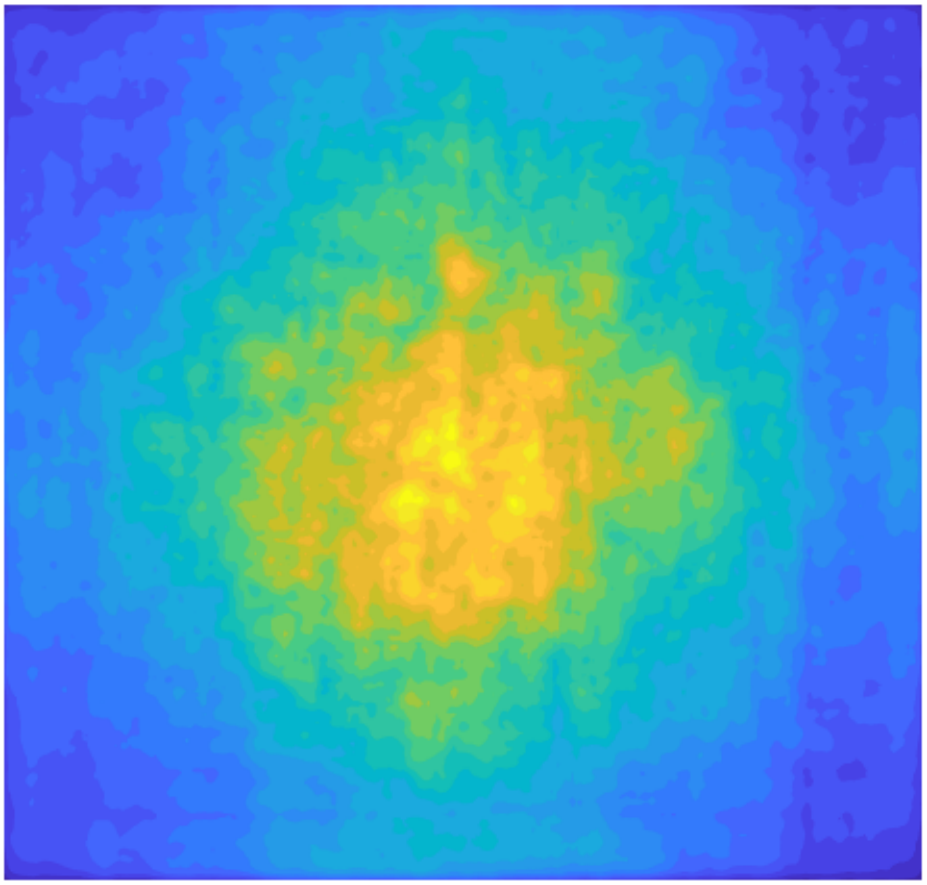}
		\subcaption{UGC}		
	\end{minipage}
	\begin{minipage}[t]{0.15\textwidth}
		\includegraphics[width=\textwidth]{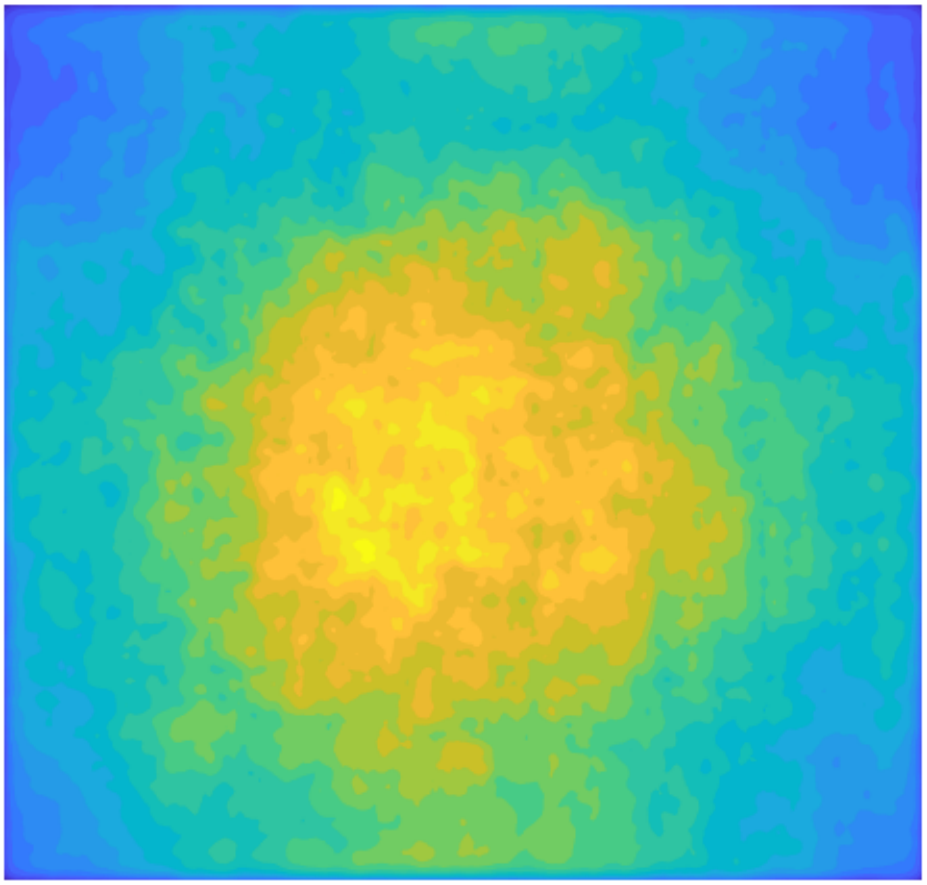}
		\subcaption{PGC}
	\end{minipage}
	\begin{minipage}[t]{0.15\textwidth}
		\includegraphics[width=\textwidth]{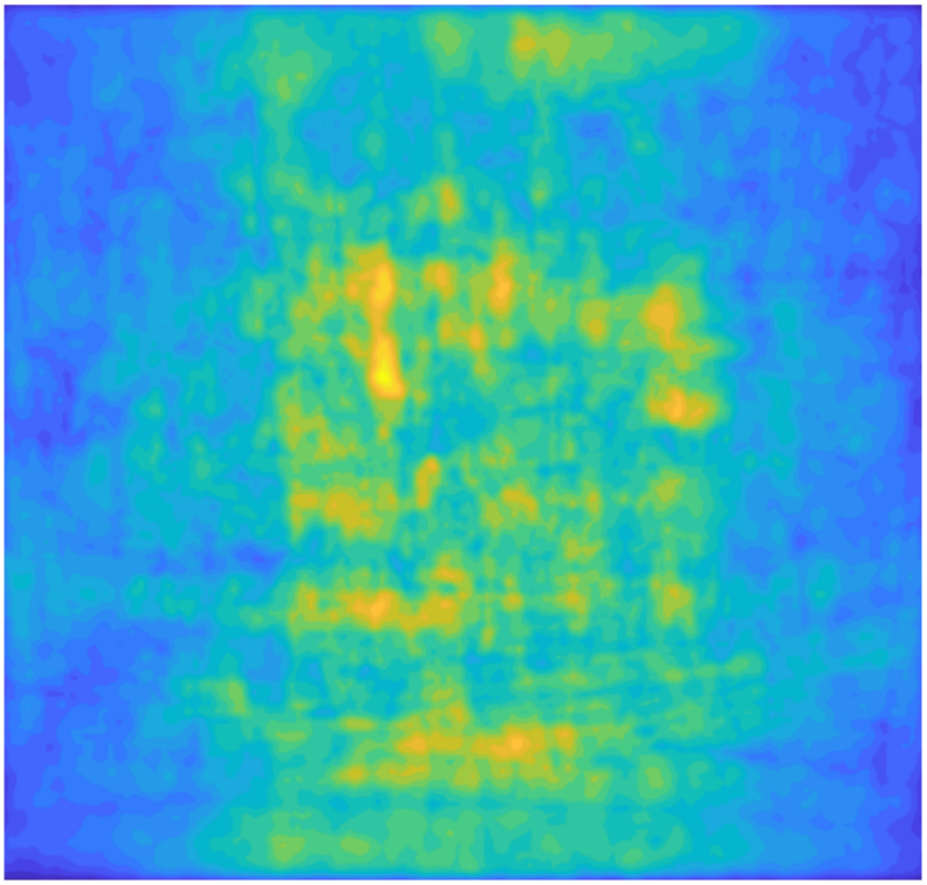}
		\subcaption{OGC}
	\end{minipage}
	\caption{The averaged visual saliency map comparison between U/P/OGC videos (1 frame computed per video). From yellow to blue indicates the visual saliency from high to low.}
	\label{fig:UPOGC-saliency}
\end{figure}

For most VQA models, the computational complexity is exponentially correlated~\cite{fastvqa} with the image size. Thus, spatial downsampling plays a key role in reducing complexity. Meanwhile, a sub-image should follow the visual saliency of HVS to reach a good VQA performance. Fig.~\ref{fig:UPOGC-saliency} shows the saliency map of UGC, PGC, and OGC through the most widely-used~\cite{SDSP} salient detection method. The result shows that for UGC, HVS tends to focus on the video's geometric center; for PGC, the size of such center doubled; for OGC, there is almost no saliency center. Therefore, when a video is more likely to be classified as UGC, the more concentrated the saliency distribution is, the more it can be downsampled in the spatial domain. Due to the commonly used deep learning model, only 7 / 8 of the image is input into the network. We assume that this sampling is the scene when $x = 0.5$ in the classification model. Therefore, the spatial sub-sampled image $img_s$ can be expressed as:
\vspace{-3mm}
\begin{equation}
img_s = img((\frac{x}{8}h:\frac{{8 - x}}{8}h),(\frac{x}{8}w:\frac{{8 - x}}{8}w))
\end{equation}
Thus, according to the characteristics of UGC, PGC, and OGC, we reduce the network input size while retaining the main area of the image.

\subsection{Temporal Domain}

For the Quality of Experience (QoE) models, existing methods
either take all video frames as input ~\cite{bentaleb2016sdndash} or intercept some
frames ~\cite{brisque} from the video, the former may result in long latency, while the latter may suffer from inadequate long-term feature representation\cite{yan2022subjective}. Therefore, to guarantee real-time prediction, we can only analyze a certain number of frames. Traditionally, the QoE model assumes that the video content of different segments contributes equally to the QoE and therefore samples each segment evenly. However, the following characteristics of video result at the end of the segment having a greater impact on QoE than at the start:
(\emph{i}) \textbf{UGC: }
 It's generally believed that for UGC videos, people will switch to the next video as soon as they are not interested in the one they are currently watching, so the content of the front part of the video is more important.
(\emph{ii}) \textbf{OGC: }
It is of higher quality and may require payment. As a result, users are more likely to watch the videos in their entirety and remember the content towards the end of the video. Therefore, the content of the later part of the video is considered to be more important.

Considering the two video characteristics mentioned above,
we conduct brisque ~\cite{brisque}, the most widely used NR-VQA database for UGC, PGC, and OGC. The overall quality of a video $Q$ and from its $n$ segments is represented as:
\vspace{-1.5mm}
\begin{equation}
    Q = \sum_{i=1}^n \omega_i Q_i
\end{equation}
\vspace{-0.5mm}
where $\omega_i$ is the weight parameter of a segment’s quality $Q_i$. Then we use Spearman Rank-order Correlation Coefficient (SRoCC) as the correlation function $S$ between QoE and the Mean Opinion Score (MOS) $M$ from the subjective assessment:
\vspace{-0.5mm}
\begin{equation}
    S(Q, M) = \sum_{i=1}^n S(\omega_i Q_i, M) = \sum_{i=1}^n \omega_i S(Q_i, M)
\end{equation}
where $fr_i$ is the number of selected frames from
the start/end of a segment. The more frames sampled, the
better the QoE model’s predictions correlated with a subjective score. As SRoCC is approximately logarithmically\cite{non-uniform} related to the sampling rate, with
a certain number of sampling frames, the specific sampling
the scheme can be transformed into an optimization problem:
\vspace{-2mm}
\begin{equation}
\label{equ:convex}
 \left\{
        \begin{array}{ll}
            fr = \sum\limits_{i=1}^n fr_i \\
            \max(\sum\limits_{i=1}^n \omega_i {\rm log}(fr_i)) 
        \end{array}
    \right.   
\end{equation}
where $fr$ is the number of frames sampled. From the LaGrange multiplier method, the derivative of the log function gives $fr$ as proportional to $w$. From (\ref{equ:convex}), the optimal sampling scheme can be derived by fully sampling one subsegment respectively, and discarding the contents of the other sub-segment to reflect the weight $\omega$ through the SRoCC between QoE and subjective score, which implies the best sampling scheme:
\vspace{-2mm}
\begin{equation}
\begin{aligned}
    \frac{fr_i}{fr_{i+1}} = \frac{\omega_i}{\omega_{i+1}} = \frac{S({\rm brisque}(seg_i), M)}{S({\rm brisque}(seg_{i+1}), M)}
\end{aligned}
\end{equation} 

where $seg_i$ is the sub-segment and QoE is predicted by ${\rm brisque}(\cdot)$ similar as Section \ref{sec:classification}. When $n\to +\infty$, the video is divided into countless segments. And we computed the ratio of $\frac{\omega_1}{\omega_n}$ in UGC\cite{database-UGC} and OGC\cite{database-HDR} databases to conduct temporal downsampling:
(\emph{i}) \textbf{UGC:}
The ratio of the weight parameters of the first and last frame $\frac{\omega_1}{\omega_n} = \frac{S({\rm brisque}(seg_1), M)}{S({\rm brisque}(seg_n), M)}=\frac{5}{3}$.
(\emph{ii}) \textbf{OGC:}
The ratio of the weight parameter is $\frac{\omega_1}{\omega_n} = \frac{5}{6}$.

Therefore, assume a linear relationship between $\omega_i$ and $t_i$, based on the previous confidence parameter $x$, we have:
\vspace{-2mm}
\begin{equation}
    \frac{\omega_1}{\omega_n} = \frac{5}{3+3x}
\end{equation}

The sampling density starts at $3+3x$ and ends at $5$. For example, while $x = 0.1$, for sampling 10 frames in a 150 frames video, the sampled frame index $fr_t(x)$ according to the above method is shown in Fig.\ref{fig:framework}. Thus, the metric of sampling different frames for different videos from UGC to OGC is realized.

\subsection{Training}

To realize low complexity while extracting the key features from the spatial downsampled sub-graph, we use Fragment Attention Network (FANet) as the backbone and two supplementary modules which have been proven effective in previous work~\cite{fastvqa}, including:
(\emph{i}) \textbf{Two separate bias tables:} One for intra-patch attention pairs and one for cross-patch attention pairs. The mechanism for the bias tables is the same as T, but they are learned separately and used for the respective attention pairs.
(\emph{ii}) \textbf{Non-linear regression:} Before performing pooling, regressing the features can avoid confusion between mini-patches with diverse qualities due to discontinuity between them. The output score $s_{pred}$ can be expressed as:
    \begin{equation}
        s_{pred} = Pool(R_{NL}(f))
    \end{equation}
    where $R_{NL}$ is non-linear layer and $f$ is the feature.
 
Overall, these modules are added to Swin-T to adapt it to image fragments and improve its performance.

\begin{table*}[t]
\centering
\small
\caption{Performance results on the UGC, PGC, and OGC databases. The best performance results are marked in {\bf\textcolor{red}{RED}}; the second SRoCC, KRoCC, and PLCC performance and the computational time under 5s are marked in {\bf\textcolor{blue}{BLUE}}.}
\label{tab:result}
\begin{tabular}{l|llll|llll|llll}
\hline
\multicolumn{1}{c|}{\multirow{2}{*}{Metric}} & \multicolumn{4}{c|}{UGC}                                                                 & \multicolumn{4}{c|}{PGC}                                                                 & \multicolumn{4}{c}{OGC}                                                                  \\ \cline{2-13} 
\multicolumn{1}{c|}{}                        & SRoCC           & KRoCC           & \multicolumn{1}{l|}{PLCC}            & Time          & SRoCC           & KRoCC           & \multicolumn{1}{l|}{PLCC}            & Time          & SRoCC           & KRoCC           & \multicolumn{1}{l|}{PLCC}            & Time          \\ \hline
Brisque\cite{brisque}                                      & 0.3820          & 0.2635          & \multicolumn{1}{l|}{0.3952}          & \textbf{\textcolor{red}{1.70}} & 0.5758          & 0.4006          & \multicolumn{1}{l|}{0.5734}          & \textbf{\textcolor{red}{2.81}} & 0.3276          & 0.2441          & \multicolumn{1}{l|}{0.3452}          & \textbf{\textcolor{blue}{4.57}} \\ \hline
Niqe\cite{B:Niqe}                                         & 0.3929          & 0.2732          & \multicolumn{1}{l|}{0.4027}          & \textbf{\textcolor{blue}{3.11}}          & 0.4709          & 0.3252          & \multicolumn{1}{l|}{0.4258}          & \textbf{\textcolor{blue}{4.89}}          & 0.3695          & 0.2770          & \multicolumn{1}{l|}{0.2869}          & 10.5          \\ \hline
Piqe\cite{B:Piqe}                                         & 0.4028          & 0.2742          & \multicolumn{1}{l|}{0.4157}          & \textbf{\textcolor{blue}{3.03}}          & 0.4329          & 0.3249          & \multicolumn{1}{l|}{0.4225}          & 5.92          & 0.6186          & 0.4563          & \multicolumn{1}{l|}{0.6376}          & 11.7          \\ \hline
Viideo\cite{B:Viideo}                                       & 0.0671          & 0.0279          & \multicolumn{1}{l|}{0.0977}          & 11.6          & 0.2431          & 0.1630          & \multicolumn{1}{l|}{0.2637}          & 14.2          & 0.1589          & 0.0980          & \multicolumn{1}{l|}{0.1555}          & 18.8          \\ \hline
V-blinds\cite{B:Vblind}                                     & 0.5855          & 0.4047          & \multicolumn{1}{l|}{0.5961}          & 106           & 0.6939          & 0.5078          & \multicolumn{1}{l|}{0.7172}          & 275           & \textbf{\textcolor{blue}{0.6695}}          & \textbf{\textcolor{blue}{0.4877}}          & \multicolumn{1}{l|}{\textbf{\textcolor{blue}{0.6508}}}          & 456           \\ \hline
ResNet50\cite{B:Resnet}                                     & 0.7183          & 0.5229          & \multicolumn{1}{l|}{0.7097}          & 5.24          & 0.6636          & 0.4786          & \multicolumn{1}{l|}{\textbf{\textcolor{blue}{0.7205}}}          & 6.31          & 0.6014          & 0.4545          & \multicolumn{1}{l|}{0.6335}          & 9.63          \\ \hline
Fast-VQA\cite{fastvqa}                                     & \textbf{\textcolor{red}{0.8477}} & \textbf{\textcolor{red}{0.7272}} & \multicolumn{1}{l|}{\textbf{\textcolor{red}{0.8136}}} & \textbf{\textcolor{blue}{4.15}}          & \textbf{\textcolor{blue}{0.7373}}          & \textbf{\textcolor{blue}{0.5426}}          & \multicolumn{1}{l|}{0.6986}          & \textbf{\textcolor{blue}{4.30}}          & 0.5030          & 0.3788          & \multicolumn{1}{l|}{0.5444}          & \textbf{\textcolor{red}{4.37}} \\ \hline
\textbf{XGC-VQA}                                     & \textbf{\textcolor{blue}{0.8245}}         & \textbf{\textcolor{blue}{0.7033}}           & \multicolumn{1}{l|}{\textbf{\textcolor{blue}{0.7805}} }          & \textbf{\textcolor{blue}{4.32}}           & \textbf{\textcolor{red}{0.7949}} & \textbf{\textcolor{red}{0.6138}} & \multicolumn{1}{l|}{\textbf{\textcolor{red}{0.7541}}} & \textbf{\textcolor{blue}{4.58}}          & \textbf{\textcolor{red}{0.7188}} & \textbf{\textcolor{red}{0.5336}} & \multicolumn{1}{l|}{\textbf{\textcolor{red}{0.6513}}} & \textbf{\textcolor{blue}{4.83}} \\ \hline
\end{tabular}
\vspace{-3mm}
\end{table*}

\section{Performance Evaluation}

\subsection{Experiment Setup}

The proposed metric is validated on the Youtube-UGC
~\cite{database-UGC}, Live-VQC (PGC) ~\cite{database-VQC}, and Live-HDR (OGC) ~\cite{database-HDR} databases. Youtube-UGC is the most widely used NR-VQA database for UGC content. Due to the recent reduction of UGC quality as Section \ref{sec:relate} mentioned, we removed some UGC with excessively high resolution; Live-VQC is a large-scale video quality assessment database which is commonly regarded as PGC\cite{database-ugcpgcogc}, whose content is better than UGC. Live-HDR is a database for HDR videos, whose processing technology of improving image brightness and contrast is in line with the high resolution and quality of OGC. Therefore Live-HDR can be used as today's OGC database.

The databases are split randomly in an
80/20 ratio for training/testing set. For the SVR and deep learning-based model, the partitioning and assessment are repeated 1,000 and 10 times for fair comparison and computational complexity, while the average result is reported as the final performance. Our metric is compared with 7 widely-used VQA metric, which shows outstanding performance in previous VQA tasks. Three major VQA types, namely handcraft~\cite{brisque,B:Niqe,B:Piqe}, Support Vector Regression (SVR)~\cite{B:Viideo,B:Vblind}, and deep learning-based\cite{B:Resnet,fastvqa} model are all included.

We use three common correlation functions, namely SRoCC, Kendall Rank-order Correlation Coefficient (KRoCC), and Pearson Linear Correlation Coefficient (PLCC), to measure how well our metric correlates with the subjective scores. The computational time is verified in seconds on an NVIDIA RTX A6000 GPU.

\begin{table}[t]
\centering
\caption{Median performance results of abanding different modules on 3 databases.}
\label{tab:ablation}
\begin{tabular}{l|lll|l}
\hline
Abandoned & SRoCC  & KRoCC  & PLCC   & Time \\ \hline
None      & \textbf{0.7794} & \textbf{0.6168} & \textbf{0.7286} & \textbf{4.58} \\ \hline
Spatial  & 0.7617 & 0.6013 & 0.7101 & 4.27 \\ \hline
Temporal   & 0.7081 & 0.5422 & 0.7093 & 4.57 \\ \hline
All       & 0.6960  & 0.5495 & 0.6855 & 4.27 \\ \hline
\end{tabular}
\vspace{-3mm}
\end{table}

\subsection{Experimental Results and Discussion}

Table.\ref{tab:result} shows the result, from which we have several useful findings. The utilization of the SVR / deep learning-based model has been found to yield superior outcomes when compared to a handcrafted model, resulting in a performance enhancement of approximately 60\%. However, this advantage is offset by a computational cost that is nearly twice as high. Under this challenge of high complexity, XGC adopts the FANet architecture similar to FastVQA, ensuring its assessment time is less than 5s for each content. Additionally, it has been observed that certain models exhibit exceptional proficiency in processing specific types of videos, but a gradual decline in coefficients has been noted across different databases. For instance, Fast-VQA demonstrates a remarkable correlation, exceeding 0.8, when evaluating UGC but produces only average results, reaching a correlation of 0.5, when processing OGC. Similarly, the V-blinds\cite{B:Vblind} has satisfying performance on OGC but gradually declined in UGC. Ultimately, the proposed XGC-VQA model demonstrates exceptional performance on all databases, while sustaining a relatively rapid processing time, particularly when processing OGC videos with a larger resolution, whose computation time greatly outperforms those of SVR-based models.

\vspace{-3mm}
\subsection{Ablation Study}

We conduct an ablation experiment to single out the
core contributors of XGC-VQA. The results are listed in Table.~\ref{tab:ablation}. The results obtained from our experiment reveal that incorporating either the time or space domain does not significantly contribute to an increase in computation time, thereby ensuring that our model operates in real-time. Conversely, omitting either domain leads to a decrease in experimental results.

\vspace{-2.2mm}
\section{Conclusions}
\vspace{-1.5mm}

Facing the challenge that the traditional VQA model cannot achieve good results on UGC, PGC, and OGC at the same time, we propose a unified VQA model: the video is classified by confidence parameter $x$ for UGC, PGC, and OGC; spatial and temporal domain sampling is done based on $x$. In addition to the pervasiveness of video content, our approach also provides real-time bitrate guidance for all types of videos on the internet today, driving the development and evolution of perception-inspired video communication.

\bibliographystyle{IEEEbib}
\vspace{-1mm}
\bibliography{icme2022template}

\begin{thebibliography}{10}

\bibitem{cisco2020cisco}
Cisco,
\newblock ``{Cisco Visual Networking Index: Forecast and Trends, 2018--2023},''
\newblock {\em White Paper}, 2020.

\bibitem{intro-performance_guidance}
S. Wang, A. Rehman, Z. Wang, S. Ma, and W. Gao,
\newblock ``Perceptual video coding based on ssim-inspired divisive
  normalization,''
\newblock {\em IEEE TIP}, 2012.

\bibitem{intro-perception}
H.F. Bermudez, J.M. Martinez-Caro, R. Sanchez-Iborra, J. Arciniegas, and M.D.
  Cano,
\newblock ``Live video-streaming evaluation using the itu-t p.1203 qoe model in
  lte networks,''
\newblock {\em Computer Networks}, 2019.

\bibitem{inspire3}
T. Zhao, Q. Liu, and C.W. Chen,
\newblock ``Qoe in video transmission: A user experience-driven strategy,''
\newblock {\em IEEE COMMUN SURV TUT}, 2017.

\bibitem{inspire2}
G. Zhai, X. Min, and N. Liu,
\newblock ``Free-energy principle inspired visual quality assessment: An
  overview,''
\newblock {\em Digital Signal Processing}, 2019.

\bibitem{intro-realtime}
J. Maisonneuve, M. Deschanel, J. Heiles, W. Li, H. Liu, R. Sharpe, and Y. Wu,
\newblock ``An overview of iptv standards development,''
\newblock {\em IEEE TBC}, 2009.

\bibitem{intro-platform}
Y. Li, S. Meng, X. Zhang, M. Wang, S. Wang, Y. Wang, and S. Ma,
\newblock ``User-generated video quality assessment: A subjective and objective
  study,''
\newblock {\em IEEE TMM}, 2023.

\bibitem{intro-PGC}
J. Kim,
\newblock ``The institutionalization of youtube: From user-generated content to
  professionally generated content,''
\newblock {\em Media, Culture \& Society}, 2012.

\bibitem{database-ugcpgcogc}
J. Xu, J. Li, X. Zhou, W. Zhou, B. Wang, and Z. Chen,
\newblock ``Perceptual quality assessment of internet videos,''
\newblock in {\em ACM MM}, 2021.

\bibitem{intro-ogcmetric1}
W. Gao, Q. Jiang, R. Wang, S. Ma, G. Li, and S. Kwong,
\newblock ``Consistent quality oriented rate control in hevc via balancing
  intra and inter frame coding,''
\newblock {\em IEEE T IND INFORM}, 2022.

\bibitem{intro-ogcmetric2}
W. Gao, S. Kwong, H. Yuan, and X. Wang,
\newblock ``Dct coefficient distribution modeling and quality dependency
  analysis based frame-level bit allocation for hevc,''
\newblock {\em IEEE TCSVT}, 2016.

\bibitem{r-spatial-map}
M. Xu, L. Jiang, X. Sun, Z. Ye, and Z. Wang,
\newblock ``Learning to detect video saliency with hevc features,''
\newblock {\em IEEE TIP}, 2016.

\bibitem{r-patch}
Z. Ying, M. Mandal, D. Ghadiyaram, and A. Bovik,
\newblock ``Patch-vq: 'patching up' the video quality problem,''
\newblock in {\em IEEE/CVF CVPR}, 2021.

\bibitem{fastvqa}
H. Wu, C. Chen, J. Hou, L. Liao, A. Wang, W. Sun, Q. Yan, and W. Lin,
\newblock ``Fast-vqa: Efficient end-to-end video quality assessment with
  fragment sampling,''
\newblock in {\em ECCV}, 2022.

\bibitem{min2020study}
X. Min, G. Zhai, J. Zhou, M.C. Farias, and A.C. Bovik,
\newblock ``Study of subjective and objective quality assessment of
  audio-visual signals,''
\newblock {\em IEEE TIP}, 2020.

\bibitem{yan2022subjective}
J. Yan, J. Li, Y. Fang, Z. Che, X. Xia, and Y. Liu,
\newblock ``Subjective and objective quality of experience of free viewpoint
  videos,''
\newblock {\em IEEE TIP}, 2022.

\bibitem{unified}
Q. Jiang, F. Shao, W. Gao, Z. Chen, G. Jiang, and Y.S. Ho,
\newblock ``Unified no-reference quality assessment of singly and multiply
  distorted stereoscopic images,''
\newblock {\em IEEE TIP}, 2019.

\bibitem{util-HDR1}
M. Dafaallah, H. Yuan, S. Jiang, and Y. Yang,
\newblock ``An attention-based network for single image hdr reconstruction,''
\newblock in {\em ISCAS}, 2022.

\bibitem{util-HDR2}
K. Zhang, Y. Fang, W. Chen, Y. Xu, and T. Zhao,
\newblock ``A display-independent quality assessment for hdr images,''
\newblock {\em IEEE SPL}, 2022.

\bibitem{sun2020dynamic}
W. Sun, X. Min, G. Zhai, K. Gu, S. Ma, and X. Yang,
\newblock ``Dynamic backlight scaling considering ambient luminance for mobile
  videos on lcd displays,''
\newblock {\em IEEE TMC}, 2020.

\bibitem{class-brightness}
S. Wang, J. Zheng, H.M. Hu, and B. Li,
\newblock ``Naturalness preserved enhancement algorithm for non-uniform
  illumination images,''
\newblock {\em IEEE TIP}, 2013.

\bibitem{database-UGC}
J.G. Yim, Y. Wang, N. Birkbeck, and B. Adsumilli,
\newblock ``Subjective quality assessment for youtube ugc dataset,''
\newblock in {\em ICIP}, 2020.

\bibitem{database-VQC}
Z. Sinno and A.C. Bovik,
\newblock ``Large-scale study of perceptual video quality,''
\newblock {\em IEEE TIP}, 2019.

\bibitem{database-HDR}
Z. Shang, J.P. Ebenezer, A.C. Bovik, Y. Wu, H. Wei, and S. Sethuraman,
\newblock ``Subjective assessment of high dynamic range videos under different
  ambient conditions,''
\newblock in {\em ICIP}, 2022.

\bibitem{class-monitor}
C.H. Chang, C.K. Liang, and Y.Y. Chuang,
\newblock ``Content-aware display adaptation and interactive editing for
  stereoscopic images,''
\newblock {\em IEEE TMM}, 2011.

\bibitem{class-JND}
S.H. Bae and M. Kim,
\newblock ``A dct-based total jnd profile for spatiotemporal and foveated
  masking effects,''
\newblock {\em IEEE TCSVT}, 2017.

\bibitem{class-quality}
W. Yang, S. Wang, Y. Fang, Y. Wang, and J. Liu,
\newblock ``Band representation-based semi-supervised low-light image
  enhancement: Bridging the gap between signal fidelity and perceptual
  quality,''
\newblock {\em IEEE TIP}, 2021.

\bibitem{yang2020fidelity}
W. Yang, S. Wang, Y. Fang, Y. Wang, and J. Liu,
\newblock ``From fidelity to perceptual quality: A semi-supervised approach for
  low-light image enhancement,''
\newblock in {\em IEEE/CVF CVPR}, 2020.

\bibitem{brisque}
A. Mittal, A.K. Moorthy, and A.C. Bovik,
\newblock ``No-reference image quality assessment in the spatial domain,''
\newblock {\em IEEE TIP}, 2012.

\bibitem{SDSP}
L. Zhang, Z. Gu, and H. Li,
\newblock ``Sdsp: A novel saliency detection method by combining simple
  priors,''
\newblock in {\em IEEE ICIP}, 2013.

\bibitem{bentaleb2016sdndash}
A. Bentaleb, A.C. Begen, and R. Zimmermann,
\newblock ``Sdndash: Improving qoe of http adaptive streaming using software
  defined networking,''
\newblock in {\em ACM MM}, 2016.

\bibitem{non-uniform}
C. Li, M. Lim, A. Bentaleb, and R. Zimmermann,
\newblock ``A real-time blind quality-of-experience assessment metric for http
  adaptive streaming,''
\newblock {\em arXiv preprint arXiv:2105.14550}, 2023.

\bibitem{B:Niqe}
A. Mittal, R. Soundararajan, and A.C. Bovik,
\newblock ``Making a “completely blind” image quality analyzer,''
\newblock {\em IEEE SPL}, 2012.

\bibitem{B:Piqe}
N. Venkatanath, D. Praneeth, M.C. Bh, S.S. Channappayya, and S.S. Medasani,
\newblock ``Blind image quality evaluation using perception based features,''
\newblock in {\em IEEE NCC}, 2015.

\bibitem{B:Viideo}
A. Mittal, M.A. Saad, and A.C. Bovik,
\newblock ``A completely blind video integrity oracle,''
\newblock {\em IEEE TIP}, 2015.

\bibitem{B:Vblind}
M.A. Saad, A.C. Bovik, and C. Charrier,
\newblock ``Blind prediction of natural video quality,''
\newblock {\em IEEE TIP}, 2014.

\bibitem{B:Resnet}
K. He, X. Zhang, S. Ren, and J. Sun,
\newblock ``Deep residual learning for image recognition,''
\newblock in {\em IEEE/CVF CVPR}, 2016.

\end{thebibliography}

\end{document}